\documentclass[conference, a4paper]{IEEEtran}

\usepackage{cite}
\usepackage{subcaption} 
\usepackage{amsmath,amssymb,amsfonts}
\usepackage{graphicx}
\usepackage{booktabs} 

\usepackage{tabularray}
\usepackage{textcomp}
\usepackage{tikz}
\usepackage{amsmath}
\usepackage{amsfonts}
\usepackage{adjustbox}
\usepackage{multirow}
\usepackage{xcolor}
\usepackage{pgfplots}
\usepackage{subcaption}
\usepackage[affil-it]{authblk} 
\usepackage{booktabs}
\usepackage{algpseudocode}
\usepackage[ruled, lined, longend, linesnumbered]{algorithm2e}
\usepackage[utf8]{inputenc}
\usepackage{tikz}

\usepackage{amssymb}
\usepackage{pifont}
\begin{document}
\title{A Life-long Learning Intrusion Detection System for 6G-Enabled IoV }

\author[2,3]{Abdelaziz Amara korba}
\author[1]{Souad Sebaa}
\author[1]{Malik Mabrouki}
\author[3]{Yacine Ghamri-Doudane}
\author[1]{Karima Benatchba}

\affil[1]{École nationale Supérieure d’Informatique, Algeria}
\affil[2]{LRS, Badji Mokhtar University of Annaba, Algeria}
\affil[3]{L3I, University of La Rochelle, France}

\maketitle
\begin{abstract}

The introduction of 6G technology into the Internet of Vehicles (IoV) promises to revolutionize connectivity with ultra-high data rates and seamless network coverage. However, this technological leap also brings significant challenges, particularly for the dynamic and diverse IoV landscape, which must meet the rigorous reliability and security requirements of 6G networks. Furthermore, integrating 6G will likely increase the IoV's susceptibility to a spectrum of emerging cyber threats. Therefore, it is crucial for security mechanisms to dynamically adapt and learn new attack patterns, keeping pace with the rapid evolution and diversification of these threats - a capability currently lacking in existing systems. This paper presents a novel intrusion detection system leveraging the paradigm of life-long (or continual) learning. Our methodology combines class-incremental learning with federated learning, an approach ideally suited to the distributed nature of the IoV. This strategy effectively harnesses the collective intelligence of Connected and Automated Vehicles
(CAVs) and edge computing capabilities to train the detection system. To the best of our knowledge, this study is the first to synergize class-incremental learning with federated learning specifically for cyber attack detection. Through comprehensive experiments on a recent network traffic dataset, our system has exhibited a robust adaptability in learning new cyber attack patterns, while effectively retaining knowledge of previously encountered ones. Additionally, it has proven to maintain high accuracy and a low false positive rate.
\end{abstract}


\section{Introduction}

As we enter the 6G era, the landscape of digital communications and interactions is undergoing a profound transformation. This evolution is particularly evident in the Internet of Vehicles (IoV) ecosystem, where the convergence of 6G technology with Connected and Automated Vehicles (CAVs) promises to significantly enhance efficiency, convenience, and user experience. However, this technological advancement is not without its risks. The adoption of 6G in the IoV potentially increases the vulnerability of CAVs to a multitude of emerging cyber threats, making security a major concern for this promising technology \cite{osorio2022towards}.

In the complex landscape of cybersecurity, Intrusion Detection Systems (IDS) hold a critical role. The transition from traditional signature-based IDS to anomaly-based systems, capitalizing on advancements in artificial intelligence (AI), represents a significant leap forward. This evolution has led to the development of IDS that can detect a wide array of cyber attacks with remarkable precision \cite{sedjelmaci2023secure,10279368, 10496859}. Predominantly, these recent AI-driven IDS rely on deep learning techniques \cite{10279368, xu2022secure}, however an inherent limitation of DL: the challenge of incremental learning from continuously evolving data streams, often described as catastrophic forgetting \cite{nature}. This challenge is particularly relevant when the system must differentiate between classes that are not observed concurrently. In essence, the IDS must be capable of dynamically adapting to new types of attacks or variations in data patterns that were absent in its initial training dataset. This capability for dynamic learning and adaptability is a critical component for IDS to remain effective in the constantly changing landscape of cyber threats.

Continual Learning, also known as life-long learning, focuses on acquiring knowledge from a continuous stream of data, aiming to expand this knowledge base without the need for retraining from scratch \cite{nature}. This area of ML has grown due to its practical benefits, such as improving medical diagnoses, advancing autonomous driving, and accurately predicting financial trends. Its growth highlights its potential to enhance AI adaptability in diverse real-world situations \cite{nature}. However, few studies \cite{ejaz2023life,prasath2022analysis} have investigated the application of continual learning for attack detection, the proposed solutions are not fully suited to the distributed and dynamic nature of the IoV. They rely on a centralized learning model that requires data collection, which raises privacy concerns.

This study aims to propose an adaptive IDS capable of learning new attack patterns while retaining those previously learned. We introduce a detection system that combines class-incremental learning (CIL) with federated learning (FL), ensuring adaptability and suitability for the distributed and dynamic environment of the IoV. To enable evolving detection of emerging cyber attacks, our approach integrates the Continual Learning with Experience And Replay (CLEAR) \cite{rolnick2019experience} method into our detection model. CLEAR effectively combines direct learning from recent data, ensuring the system remains adaptable, with indirect learning from historical data, enhancing its stability. We train this detection model using FL across local datasets from participating CAVs. In this setup, Multi-access edge computing (MEC) servers play a pivotal role in coordinating the training process by acting as parameter servers for aggregating FL model updates. The effectiveness of the proposed solution is evaluated using the 5G-NIDD \cite{samarakoon20225g} dataset, which contains real-world 5G network traffic traces. To mimic real-world conditions as closely as possible, we initially train the model on a dataset containing a single type of attack mixed with benign traffic. We then progressively introduce samples of various attack types, each time incorporating a new type. The results have demonstrably proven the IDS's ability to learn new patterns while maintaining high accuracy and a very low False Positive Rate (FPR). 

The remainder of this paper is organized as follows. Section~\ref{RT} describes related work. The design of our scheme is presented in Section~\ref{SOL}. Section~\ref{SIM} depicts the performance evaluation results, and finally, Section~\ref{CON} concludes the paper.


\section{Related Work} \label{RT}
Recent studies have increasingly focused on enhancing security within Internet of Vehicles (IoV) networks. Abdelwahab et al. \cite{boualouache20235g} conducted a comprehensive study on cybersecurity challenges within the 5G Vehicle-to-Everything (5G-V2X) networks, addressing issues across 5G infrastructure and specific use cases, among other security concerns. Osorio et al. \cite{osorio2022towards} provided a thorough analysis of security and privacy in the 6G-enabled IoV. They explored how key technological enablers such as network softwarization, blockchain, and AI/ML enhance secure communication. 

Considering Artificial Intelligence’s role, several studies have harnessed ML/DL to detect and mitigate cyber threats in IoV networks. Rahal et al. \cite{rahal2022antibotv} delved into botnet threats within IoV networks. They simulated botnet traffic to generate a dataset, subsequently applying various machine learning techniques to detect and analyze botnet traffic. To address the challenges associated with zero-day attack detection in IoV networks, in our previous work \cite{10279368}, we proposed a federated learning (FL)-based Intrusion Detection System (IDS) utilizing a deep autoencoder. To overcome the limitations of N-day attack recognition and the constraints associated with centralized FL, we extended this work \cite{10496859} by leveraging the paradigm of open-set learning with blockchain-enabled FL.

To address the challenges associated with deploying machine learning in 6G-enabled IoV, Hoang et al. \cite{xu2022secure} proposed a secure and reliable integration of Transfer Learning into the 6G-enabled IoV framework. Addressing security threats in 6G IoV networks, Sedjlmaci et al. \cite{sedjelmaci2023secure} developed a collaborative cybersecurity framework based on a multi-level FL algorithm and Stackelberg security games. In a similar vein, Zhang et al. \cite{zhang2021many} designed a sophisticated weight-based ensemble ML algorithm, optimized with many-objective techniques, for identifying anomalies in vehicular Controller Area Network (CAN) bus systems, thereby aligning with the high security demands of 6G networks. However, while these studies propose IDS solutions suitable for 6G-enabled IoV, they do not fully address the adaptability challenges in a such dynamic and constantly evolving environments where cyber threats are in continuous flux, with new ones emerging. A notable limitation of DL-based IDS is their struggle with incremental learning from continuously evolving data streams, a problem often termed as catastrophic forgetting \cite{nature}.

To tackle the challenge of adapting IDS to new attack patterns and addressing catastrophic forgetting in DL-based IDS, the study by Prasath et al. \cite{prasath2022analysis} examines the effectiveness of continual learning models for the incremental learning of novel attack patterns. This research involved both experimental and analytical studies, focusing on three key continual learning methods: learning without forgetting, experience replay, and dark experience replay. Additionally, another significant effort in employing continual learning is the work by Ejaz et al. \cite{ejaz2023life}, which investigates the use of continual learning techniques for consistent phishing detection over time. This study trained a vanilla neural network (VNN) model with deep feature embedding of HTML content in a continual learning setup. The results indicate that continual learning algorithms effectively maintain accuracy over time, albeit with slight performance decline. However, these solutions predominantly utilize centralized learning models, which may not align with the specific demands of the IoV environment. These models often require high bandwidth and result in increased latency. Moreover, there are privacy concerns, as sensitive data might be compromised during transmission to central locations.

To align with the specific demands of the IoV environment, recent studies \cite{rani2023federated, vinita2023federated} have proposed Federated Learning (FL)-based IDS for misbehavior detection in 5G and 6G-enabled IoV. Our solution uniquely combines FL and Continual Learning (CL) to ensure both adaptability and suitability within the distributed and dynamic environment of the IoV.



\section{Proposed Solution}\label{SOL}
\begin{figure*}[ht!]
    \centering
    \includegraphics[scale=0.4]{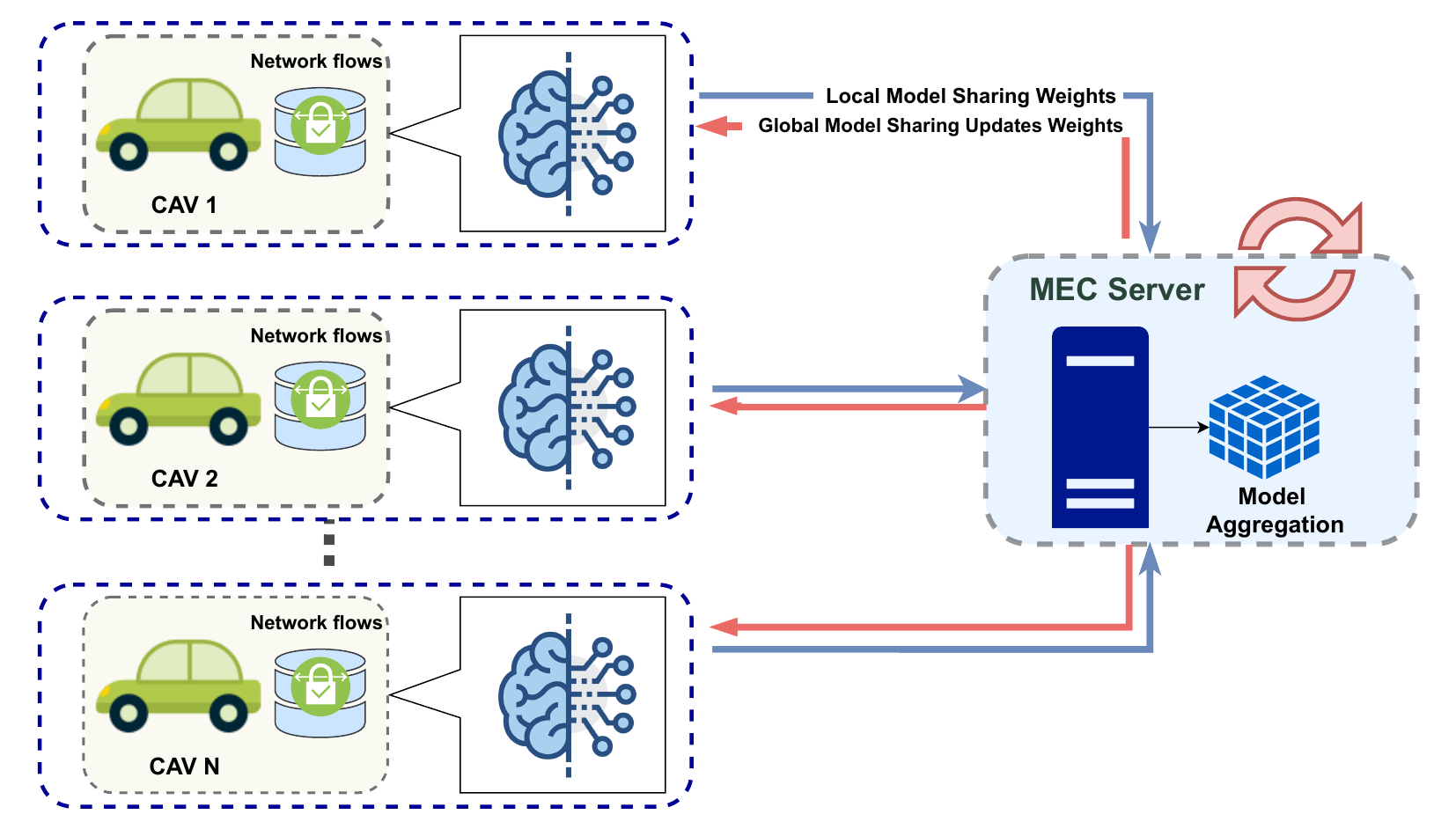}
    \caption{MEC-Enabled CIL Federated learning architecture}
    \label{fig1}
\end{figure*}

The proposed Intrusion Detection System (IDS) is specifically designed for deployment in Connected and Automated Vehicles (CAVs), where it actively monitors network traffic. Our detection model uses class-incremental learning to adapt to evolving attack patterns. Additionally, federated learning is employed to facilitate distributed and privacy-preserving training. In our system architecture, the Multi-access Edge Computing (MEC) server functions as the parameter server, with the CAVs acting as client nodes, as depicted in Figure~\ref{fig1}. We detail the development process of the IDS in the following sections.

\subsection{Network Traffic Processing} \label{sec:fe}
The traffic collector module of the IDS converts raw captured packets into flows, thereby enhancing the understanding of traffic patterns. To characterize each flow, a set of features is calculated based on packet header analysis, which includes data from both the network and transport layers of the network architecture. These network features can be categorized into four main types: Packet-Based features focus on metrics such as packet counts and transmission rates, providing insights into traffic volume and flow patterns. Byte-Based features analyze the volume of transmitted data to assess network load and bandwidth usage. Time-Based features capture the temporal characteristics of traffic, including flow durations and intervals between transmissions, aiding in the identification of activity timing and anomaly detection. Protocol-Based features leverage specific protocol information to distinguish between different types of network traffic. The resulting dataset of network flows is then utilized by the CAV for local training of the detection model.

\subsection{Federated Attack Incremental Learning}
To enable continual learning of the detection model, we use the Continual Learning with Experience And Replay (CLEAR) \cite{rolnick2019experience} method for our detection model. This methods fuses direct learning from new experiences to preserve the system's adaptability with indirect learning from past experiences to bolster stability. Moreover, CLEAR strengthens the system's consistency by incorporating behavioral cloning, aligning the current operational guidelines with prior iterations. CLEAR is underpinned by an actor-critic training regime that leverages both fresh and historical experiences. Formally, the network parameters are denoted by \(\theta\), with \(\pi_\theta\) indicating the network's current policy over actions \(a\), and \(h_s\) representing the hidden state of the network at time \(s\). The policy generating the observed experience is represented by \(\mu\). The \(V\)-Trace target \(v_s\) for this method is defined as follows \cite{rolnick2019experience}:
\[
v_s := V(h_s) + \sum_{t=s}^{s+n-1} \gamma^{t-s} \left( \prod_{i=s}^{t-1} c_i \right) \delta V_t,
\]

where \(\delta V_t := \rho_t (r_t + \gamma V(h_{t+1}) - V(h_t))\) for truncated importance sampling weights \(c_i := \min\left(\overline{c}, \frac{\pi_\theta(a_i | h_i)}{\mu(a_i | h_i)}\right)\), and \(\rho_t := \min\left(\overline{\rho}, \frac{\pi_\theta(a_t | h_t)}{\mu(a_t | h_t)}\right)\) (with \(\overline{c}\) and \(\overline{\rho}\) constants). The policy gradient loss is:
\[
L_{\text{policy-gradient}} := -\rho_s \log \pi_\theta(a_s|h_s) (r_s + \gamma v_{s+1} - V(h_s)).
\]

The loss functions \(L_{\text{policy-gradient}}\), \(L_{\text{value}}\), and \(L_{\text{entropy}}\) are utilized for both new and replay experiences. Additionally, \(L_{\text{policy-cloning}}\) and \(L_{\text{value-cloning}}\) are incorporated exclusively for replay experiences.

In replay experiences, additional loss terms are incorporated to enable behavioral cloning between the network's current state and its historical counterparts. This is aimed at preventing deviations in the network's output on replayed tasks during the learning of new tasks. The methodology includes penalizing (1) the Kullback-Leibler (KL) divergence, which assesses the disparity between the historical and present policy distributions, and (2) the L2 norm, reflecting the variance between historical and current value functions. Technically, this leads to the integration of specific loss functions \cite{rolnick2019experience}:
\[
L_{\text{policy-cloning}} := \sum_a \mu(a|h_s) \log \frac{\mu(a|h_s)}{\pi_\theta(a|h_s)},
\]
\[
L_{\text{value-cloning}} := \| V(\theta_{h_s}) - V_{\text{replay}}(h_s) \|^2.
\]


\begin{algorithm}[ht!]
    \caption{Federated Averaging Algorithm}
    \label{FL}
    \textbf{Variables}: \textit{K}: index of clients, \textit{B}: local batch size, \textit{E}: number of local epochs, \textit{$\eta$} : learning rate   \\

     Initialize ${w}_{0}$ \\
     \For{ each round $t \in \{1,...,N\}$}{
        $m\leftarrow$ max(C. K, 1) \\
        $S_{t}\leftarrow$ (random set of m CAVs) \\
        \For{each CAV $k \in S_{t} $ in parallel}{
            $\beta_k \leftarrow$ (split $\mathit{P_{k}}$ into batch of size $\mathit{B}$) \\
            \For{each local epoch $i \in \{1,...,E\}$}{
                \For{batch $b \in \beta_k$}{
                    $w_k\leftarrow \textit{w}_k - \eta \nabla l(w_k;b)$ \\
                }
            }
            $w_{t+1}^{k}\leftarrow w_{k}$ \\ 
        }
        $w_{t+1}\leftarrow \sum_{k=1}^{n}\frac{n_{k}}{n} w_{t+1}^{k}$ \\ 
    }
\end{algorithm}

The MEC nodes orchestrate the federated training process of the detection model by playing the role of parameters servers for model updates aggregation. First, the MEC server initializes the learning parameters of the detection model (number of layers, number of neurons, activation functions,learning rate, etc.). It then shares such parameters with the participant CAVs. Each CAV trains the detection model on its local dataset. Once done, each CAV sends its local model's parameters (weights) to the MEC server. The latter aggregates the received local models to generate a global learning model, before sending it back to the involved CAVs, in order to initiate a new training round. Algorithm \ref{FL} illustrates the main steps performed by both the MEC server and each participating CAV. In our study, the federated learning problem across multiple CAVs is formulated as a federated optimization problem and resolved using the FedAvg algorithm \cite{FedAvg}. Indeed, using its local data, each CAV calculates the average gradient on top of the model $w$ for a corresponding training round $r$. Thereafter, each CAV performs a local gradient descent on the currently used model with its own data. On the other hand, the MEC server aggregates these local updates and transfers back the global model to the CAV collaborators. This process is repeated during a number of rounds, defined initially by the MEC server.

\section{Performance Evaluation}\label{SIM}
\subsection{Dataset}
We evaluated our system using the 5G Network-Intrusion Detection and Defense (5G-NIDD) \cite{samarakoon20225g} dataset. This dataset was selected for its recency, the variety of attack types it includes, and particularly because it contains real 5G network traffic. Ideally, we would have preferred to use a dataset with 6G traffic, but to our knowledge, such a dataset is not yet available. The dataset includes examples of Denial of Service (DoS) attacks, such as ICMP Flood, UDP Flood, SYN Flood, HTTP Flood, and Slowrate DoS, as well as port scans, including SYN Scan, TCP Connect Scan, and UDP Scan. Following a comprehensive data preprocessing that involved cleaning, normalization, and feature engineering, the final dataset was refined to include 81 features. The sample distribution of network traffic per attack is presented in Table~\ref{tab:dataset}.

\begin{figure*}[ht!]
    \centering
    \includegraphics[scale=0.5]{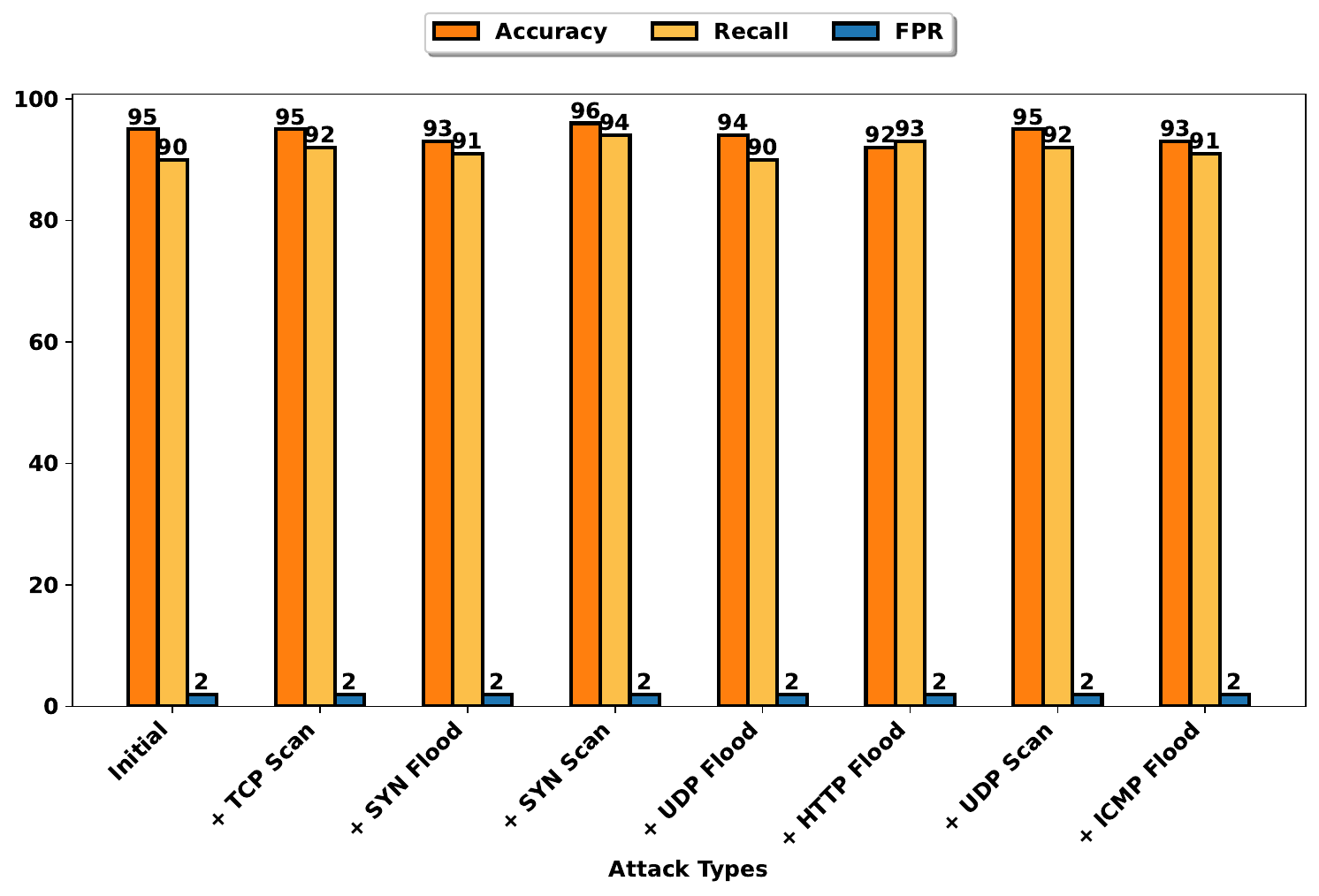}
    \caption{Evaluation of Performance Throughout the Incremental Learning Process}
    \label{fig:central}
\end{figure*}
\begin{figure*}[h!]
    \centering
    \begin{subfigure}{.4\linewidth}
        \centering
        \includegraphics[width=\linewidth]{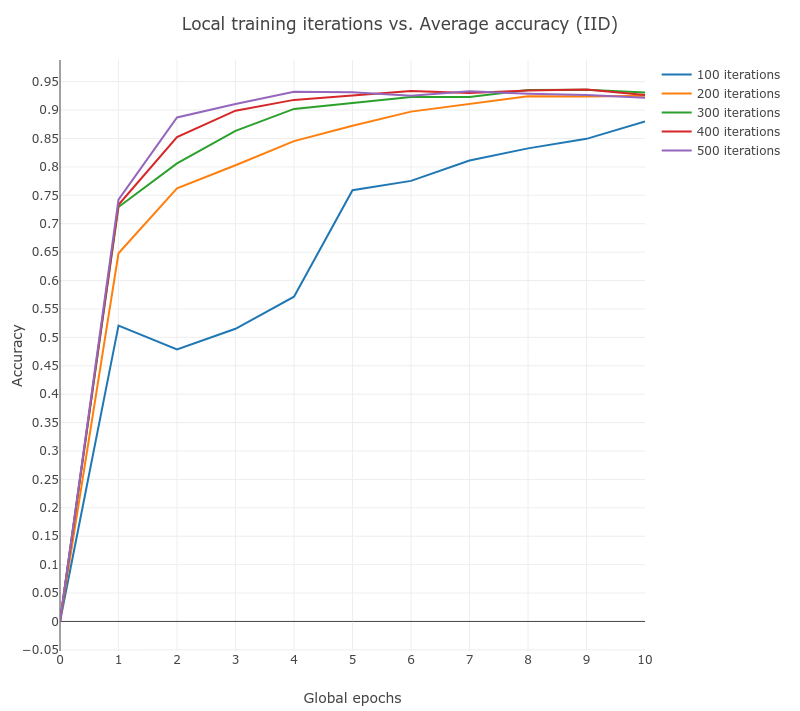} 
        \caption{\textcolor{blue}{}}
        \label{fig10a}
    \end{subfigure}
    \begin{subfigure}{.4\linewidth}
        \centering
        \includegraphics[width=\linewidth]{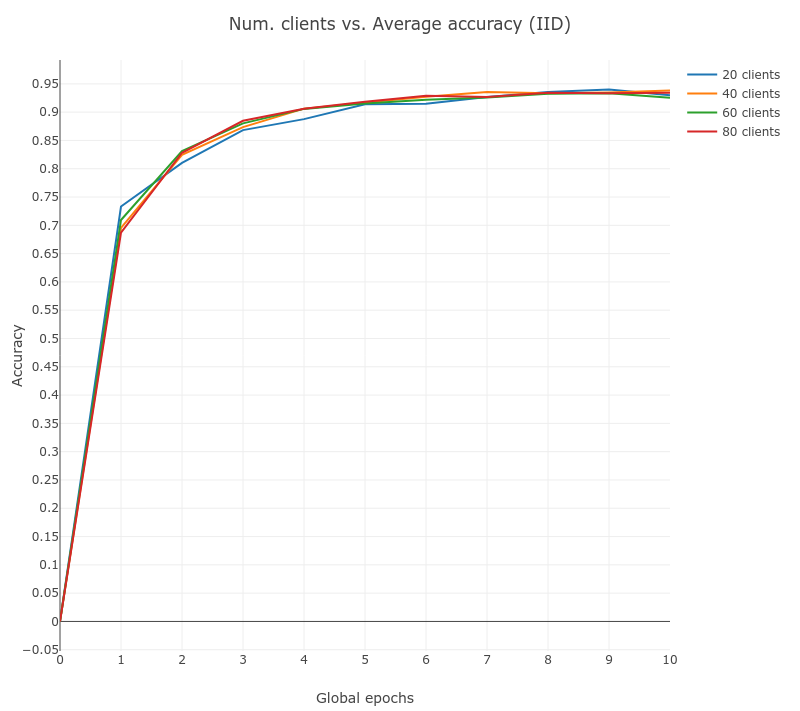} 
        \caption{\textcolor{blue}{}}
        \label{fig10c}  
    \end{subfigure}
    \caption{Accuracy vs. Nb. clients \& Nb. Iterations }
    \label{fig:modShare-acc}
\end{figure*}

\subsection{Experimental results}
We first assess our detection model's performance within a centralized training setup. Subsequently, we proceed to evaluate its accuracy in a federated training setup. To ensure a validation as realistic and closely aligned with real-world scenarios as possible, we begin by training the model on a dataset that includes only one type of attack along with benign traffic. Subsequently, we sequentially introduce samples of different attack types, each time integrating a new type. The order of the attacks' introduction is completely random. The model evaluation was based on the three metrics detailed below: 

\begin{itemize}
    \item Accuracy: the proportion of correctly identified samples (both true positives and true negatives) to the total number of observations.
    \begin{equation}
        Accuracy = \frac{TP + TN}{TP + TN + FP + FN}
    \end{equation}

    \item Recall: measures the proportion of actual positives that are correctly identified.
    \begin{equation}
        Recall = \frac{TP}{TP + FN}
    \end{equation}

    \item FPR (False Positive Rate): Measures the proportion of actual negatives that are incorrectly identified as positives.
    \begin{equation}
        FPR = \frac{FP}{FP + TN}
    \end{equation}

    \item F1-Score: The harmonic mean of Precision and Recall.
    \begin{equation}
        F1\_Score = 2 \times \frac{Precision \times Recall}{Precision + Recall}
    \end{equation}
\end{itemize}

TP, TN, FP, and FN denote true positive, true negative,
false positive, and false negative, respectively. 

\begin{table}[]
\centering
\color{black}
\caption{Dataset samples distribution}
\label{tab:dataset}
\begin{tabular}{@{}lll@{}}
\toprule
Attack      & Absolute count & Fraction (\%) \\ \midrule
Benign      & 477737         & 39.29  \\
UDPFlood    & 457340         & 37.61   \\
HTTPFlood   & 140812         & 11.58   \\
SlowrateDoS & 73124          & 6.01    \\
TCPConScan  & 20052          & 1.65    \\
SYNScan     & 20043          & 1.65    \\
UDPScan     & 15906          & 1.31    \\
SYNFlood    & 9721           & 0.80    \\
ICMPFlood   & 1155           & 0.09    \\ \bottomrule
\end{tabular}
\end{table}

\begin{table*}[h]
\centering
\resizebox{0.9\textwidth}{!}{%
\begin{tabular}{rrrrrrrrr}
\toprule
It. locales & Multiclass Acc. & Macro Recall & Weighted Recall & Binary Acc. & Binary FPR & Binary Precision & Binary Recall & Binary F1-Score \\
\midrule
100 & 0.839 & 0.642 & 0.839 & 0.960 & 0.025 & 0.969 & 0.941 & 0.955 \\
200 & 0.919 & 0.854 & 0.919 & 0.980 & 0.014 & 0.983 & 0.973 & 0.978 \\
300 & 0.931 & 0.874 & 0.931 & 0.985 & 0.020 & 0.976 & 0.991 & 0.984 \\
400 & 0.926 & 0.875 & 0.926 & 0.980 & 0.031 & 0.964 & 0.993 & 0.978 \\
500 & 0.921 & 0.874 & 0.921 & 0.975 & 0.040 & 0.954 & 0.993 & 0.973 \\
\bottomrule
\end{tabular}
}
\caption{No. of Local Iterations vs. Predictive performances}
\label{tab:fl-local-it-non-iid-metrics}
\end{table*}

\begin{table*}[h]
\centering
\resizebox{0.9\textwidth}{!}{%
\begin{tabular}{rrrrrrrrr}
\toprule
No. of Clients & Multiclass Acc. & Macro Recall & Weighted Recall & Binary Acc. & Binary FPR & Binary Precision & Binary Recall & Binary F1-Score \\
\midrule
20 & 0.928 & 0.874 & 0.928 & 0.984 & 0.024 & 0.972 & 0.993 & 0.982 \\
40 & 0.933 & 0.874 & 0.933 & 0.987 & 0.016 & 0.980 & 0.991 & 0.986 \\
60 & 0.923 & 0.870 & 0.923 & 0.979 & 0.033 & 0.961 & 0.993 & 0.977 \\
80 & 0.927 & 0.872 & 0.927 & 0.984 & 0.024 & 0.972 & 0.993 & 0.982 \\
\bottomrule
\end{tabular}
}
\caption{Nb. of Clients vs.  Predictive performances}
\label{tab:fl-cl-iid-metrics}
\end{table*}

\subsubsection{Centralized training}
We implemented and evaluated our proposed system within the Google Colab cloud environment, utilizing the Pytorch package to implement both local and federated learning models. The core of our system is a Multilayer Perceptron (MLP) model, intricately structured with three hidden layers. Each layer comprises 300 neurons, activated by the ReLU (Rectified Linear Unit) function. Throughout its training phase, it undergoes 1000 iterations. Additionally, we incorporated a buffer memory capable of storing up to 100 samples. This memory plays a crucial role in our implementation of CLEAR method. By preserving previously observed samples, it significantly mitigates the issue of catastrophic forgetting.

Figure \ref{fig:central} illustrates the performance dynamics as new types of attacks were incrementally introduced to the system. Initially, the system's handling of benign traffic and Lowrate DoS attack was exemplary, achieving 100\% accuracy, a 0\% FPR, and a detection rate of over 99\%. Despite a slight decline in performance following the addition of UDP Flood and HTTP Flood attacks, the model successfully maintained high levels of performance. Upon the final integration of all attack types, the detection model sustained robust performance, with both accuracy and recall exceeding 92\%, while maintaining a very low FPR of 2\%.

\subsubsection{Federated training}
In our federated training approach, we implemented an independent and identically distributed (IID) sampling setup. We followed the same testing strategy as our previous experiment, which involves both sequential and random observations of various attack types. In our initial setup, we considered 10 clients and conducted training over 10 rounds. We experimented with varying numbers of local iterations to assess the model's convergence. As shown in Figure \ref{fig10a}, it is evident that the model converges after 300 local iterations (across 10 FL rounds). This observation is corroborated by the swift learning rate and consistently high accuracy depicted in the plot.

For a comprehensive evaluation, we assessed the system in both binary classification (benign, malicious) and multi-class classification scenarios. In the multi-class context, we considered both Macro Recall and Weighted Recall. Macro Recall measures the average performance across all classes, treating each class equally. Weighted Recall, on the other hand, accounts for the frequency of each class, providing a more realistic evaluation in scenarios where some attacks are more prevalent than others. These metrics offer a well-rounded assessment, considering both majority and minority attacks, providing a comprehensive perspective on the model's capabilities and opportunities for improvement. As illustrated in Table \ref{tab:fl-local-it-non-iid-metrics}, the detection model shows very good performances with a high weighted recall of 0.931, showcasing its strong performance in identifying majority classes. While the macro recall is at 0.874, this still represents a commendable performance, especially in the context of less frequent classes.

To assess the system's scalability, we experimented with varying the numbers of FL CAVs. Figure \ref{fig10c} indicates strong scalability in the federated learning process. Despite increasing the number of clients per MEC node, from 20 to 80, there is no significant variance in accuracy, suggesting that the system can handle scaling horizontally—adding more clients—without a loss in performance. The quick convergence and stable high accuracy across all client configurations demonstrate the system's robustness and effectiveness in a distributed learning context.  Table \ref{tab:fl-cl-iid-metrics} demonstrates that the detection model maintains stable and high performance in both multiclass and binary classifications across varying numbers of clients. The model achieves high accuracy and precision, indicating its robustness. While the macro recall indicates a slightly lower performance in identifying less frequent attack types, the high F1-scores across all client groups suggest a well-balanced model. The consistent performance, regardless of the number of clients, affirms the model's scalability and effectiveness.

\section{Conclusion} \label{CON}
This study introduced a new adaptive IDS designed for the constantly evolving IoV security environment, in anticipation of the upcoming 6G technology shift. By integrating class-incremental and federated learning, which suited the IoV's distributed structure, the system achieved high accuracy and a low false positive rate, as demonstrated in our tests using a recent dataset. These results emphasized the system's flexibility and marked a step forward in AI-driven cybersecurity for vehicular networks. Our efforts provided a foundational step for enhancing IoV security against the new wave of cyber threats expected with 6G advancements.






\section*{Acknowledgment}

This work was supported by the 5G-INSIGHT bilateral project (ID: 14891397) / (ANR-20-CE25-0015-16), funded by the Luxembourg National Research Fund (FNR), and by the French National Research Agency (ANR).


\bibliographystyle{unsrt}
\bibliography{ref}

\end{document}